\documentclass[%
preprint,
showpacs,
preprintnumbers,
 amsmath,
 amssymb,
 aps,
pra,
longbibliography,
]{revtex4-1}

\usepackage{graphicx}
\usepackage{pslatex}

\usepackage[left]{eurosym}

\begin{document}

\title{On the plasticity of nonlocal quantum correlations}

\author{Karl Svozil}
\email{svozil@tuwien.ac.at}
\homepage{http://tph.tuwien.ac.at/~svozil}
\affiliation{Institute for Theoretical Physics, Vienna University of Technology,  \\
Wiedner Hauptstra\ss e 8-10/136, A-1040 Vienna, Austria}

\begin{abstract}
The quantum correlations of two or more entangled particles present the possibility of stronger-than-classical outcome coincidences.  We investigate two-partite correlations of spin one, three-half and higher quanta in a state satisfying a uniqueness property in the sense that knowledge of an outcome of one particle observable entails the certainty that, if this observable were measured on the other particle(s) as well, the outcome of the measurement would be a unique function of the outcome of the measurement performed. We also investigate correlations of four spin one-half particles.
\end{abstract}

\pacs{03.67.Hk,03.65.Ud}
\keywords{Quantum information, quantum communication, singlet states, group theory, entanglement, quantum nonlocality}

\maketitle



\section{Introduction}

The possibility of a peculiar and ``mind-boggling'' type of connectedness
between two or more spatially separated particles beyond classical correlations
surprised the quantum pioneers in their early exploration of quanta.
Already Schr\"odinger
noted that a state of several quantized particles or quanta
could be {\em entangled} (in Schr\"odinger's own German terminology {\em ``verschr\"ankt''})
in the sense that it cannot be represented as the product of states of the isolated, individual quanta,
but is rather defined by the {\em joint} or {\em relative} properties of the quanta involved~\cite{zeil-99,zeil-Zuk-bruk-01}.
Typical examples of such joint properties of entangled states are the propositions,
``when measured along two or more different directions, two spin one-half particles have opposite spin''
(defining the Bell singlet state),
or ``when measured along a particular direction, three spin one-half particles have identical spin''
(one of the three defining properties of the Greenberger-Horne-Zeilinger-Mermin state).

With respect to the outcome of certain measurements on the individual particles in an entangled state,
the observation of stronger-than-classical correlations for nonlocal, i.e., spatially and even causally separated, quanta
in  ``delayed choice'' measurements has been experimentally verified~\cite{wjswz-98}.
A typical phenomenological criterion of such correlations it the {\em increased} of {\em decreased} frequency of the occurrence of certain coincidences of outcomes,
such as the more- or less-often-than-classically expected recordings of joint spin up and down measurements labeled by ``$++$,'' ``$+-$,'' ``$-+$'' or ``$--$,'' respectively.

Stated pointedly, the ``magic'' behind the quantum correlations as compared to classical correlations resides in the fact that,
for almost all measurement directions (despite collinear or orthogonal ones), an observer ``Alice,''
when recording some outcome of a measurement, can be sure that her partner ``Bob,''
although spatially and causally disconnected from her, is either more or less likely to record a particular measurement outcome on his side.
However, because of the randomness  and uncontrollability of the individual events, and
because of the no-cloning theorem,
no classically useful information can be transferred from Alice to Bob, or {\it vice versa:}
The parameter independence and outcome dependence of otherwise random events ensures that
the nonlocal correlations among quanta cannot be directly used to communicate classical information.
The correlations of the joint outcomes on Alice's and Bob's sides can only be  verified by collecting all the different outcomes {\it ex post facto,}
recombining joint events one-by one.
Nevertheless, there are hopes and visions to utilize nonlocal quantum correlations for a wide range of explanations and
applications; for instance in quantum information theory~\cite{bruk-06} and life sciences~\cite{sum-05}.

In what follows a few known and novel quantum correlations will be systematically enumerated.
We shall derive the correlations between two and four two-state particles in singlet states.
We also derive the correlations of two three-, four- and general $d$-state particles in a singlet state.
In doing so we attempt to ``sharpen'' the nonclassical behavior beyond the standard quantum correlations.

\section{Two particle correlations}

In what follows, consider two particles or quanta. On each one of the two quanta, certain measurements
(such as the spin state or polarization) of
(dichotomic) observables
$O({ a})$ and
$O({ b})$
along the directions $a$ and $b$, respectively, are performed.
The individual outcomes are
encoded or labeled by the symbols ``$-$'' and  ``$+$,'' or values ``-1'' and ``+1'' are recorded along
the directions ${ a}$ for the first particle, and  ${ b}$ for the second particle, respectively.
A two-particle correlation function $E(a,b )$
is defined by averaging over the product of the outcomes $O({ a})_i, O({ b} )_i\in \{-1,1\}$
in the $i$th experiment for a total of $N$ experiments; i.e.,
\begin{equation}
E(a,b )={1\over N}\sum_{i=1}^N O({ a})_i O({ b})_i.
\end{equation}

Quantum mechanically, we shall follow a standard procedure for obtaining the probabilities upon which the correlation coefficients are based.
We shall start from the angular momentum operators, as for instance defined in Schiff's {\em ``Quantum Mechanics''}~\cite[Chap.~VI, Sec.24]{schiff-55}
in arbitrary directions, given by the spherical angular momentum co-ordinates $\theta$ and $\varphi$, as defined above.
Then, the projection operators corresponding to the eigenstates associated with the different eigenvalues are derived
from the dyadic (tensor) product of the normalized eigenvectors.
In Hilbert space based  quantum logic, every projector corresponds to
a proposition that the system is in a state corresponding to that observable.
The quantum probabilities associated with these eigenstates are derived from the Born rule, assuming singlet states for the physical reasons discussed above.
These probabilities contribute to the  correlation coefficients.


\subsection{Three-state particles}

\subsubsection*{Observables}
The angular momentum operator in arbitrary direction $\theta$, $\varphi$ is given by its spectral decomposition
\begin{equation}
S_1 (\theta ,\varphi)
= -F_{-}(\theta ,\varphi)+0\cdot F_0(\theta ,\varphi) +F_{+}(\theta ,\varphi),
\label{e-2009-gtq-s3}
\end{equation}
where $F_-,F_0,F_+$ are the orthogonal projectors associated with the eigenstates of $S_1 (\theta ,\varphi)$.
The generalized one-particle observable with the previous outcomes of spin state measurements ``coded''
into the map $
-1 \mapsto  \lambda_{-}$,
$0 \mapsto   \lambda_{0}$,
$+1 \mapsto   \lambda_{+}$
can be written as
$
R_1(\theta ,\varphi) = \lambda_{-} F_{-}(\theta ,\varphi) + \lambda_{0} F_0(\theta ,\varphi) +  \lambda_{+} F_{+}(\theta ,\varphi)
$.

For the sake of an operationalization of the 117 contexts contained in their proof,
Kochen and Specker~\cite{kochen1} introduced an observable based on spin one
with degenerate eigenvalues corresponding to
$\lambda_+ = \lambda_- = 1$ and $\lambda_0 = 0$,
or its ``inverted'' form $\lambda_+ = \lambda_- = 0$ and $\lambda_0 = 1$.
The corresponding correlation functions will be discussed below.

\subsubsection*{Singlet state}

Consider the two spin-one particle singlet state
$
\vert \Psi_{3,2,1} \rangle  =  \frac{1}{\sqrt{3}}\left(-|00\rangle + |-+\rangle + |+-\rangle \right)
$.
Its vector space representation can be explicitly enumerated by taking the direction $\theta =\varphi =0$ and recalling that
$\vert +\rangle \equiv (1,0,0)$,
$\vert 0\rangle \equiv (0,1,0)$, and
$\vert -\rangle \equiv (0,0,1)$; i.e.,
$
\vert \Psi_{3,2,1} \rangle  \equiv  \frac{1}{\sqrt{3}}\left(0,0,1,0,-1,0,1,0,0 \right)
$.

\subsubsection*{Results}

The  general computation of the quantum correlation coefficient yields
\begin{equation}
\begin{array}{ll}
&E_{{ \Psi_{3,2,1}}\,\lambda_- \lambda_0 \lambda_+ } ({\hat \theta},{\hat \varphi} )= {\rm Tr}\left[\rho_{ \Psi_{3,2,1}} \cdot R_{1 1} \left({\hat \theta},{\hat \varphi} \right)\right] =\\
& \quad   =  \frac{1}{192} \left\{24 \lambda_0^2 + 40 \lambda_0 \left(\lambda_- + \lambda_+\right) + 22 \left(\lambda_- + \lambda_+\right)^2 -
   32 \left(\lambda_- - \lambda_+\right)^2 \cos \theta_1  \cos \theta_2  +   \right.                                                                                     \\
& \qquad \; + 2 \left(-2 \lambda_0 + \lambda_- + \lambda_+\right)^2 \cos\left(2  \theta_2 \right) \left[\left(3 + \cos\left(2 \left( \varphi_1  -
  \varphi_2 \right)\right)\right) \cos\left(2  \theta_1 \right) + 2 \sin\left( \varphi_1  -  \varphi_2 \right)^2\right] +                                          \\
& \qquad \; +  2 \left(-2 \lambda_0 + \lambda_- + \lambda_+\right)^2 \left[\cos\left(2 \left( \varphi_1  -  \varphi_2 \right)\right) +
     2 \cos\left(2  \theta_1 \right) \sin\left( \varphi_1  -  \varphi_2 \right)^2\right] -                                                                         \\
& \qquad \; -32 \left(\lambda_- - \lambda_+\right)^2 \cos\left( \varphi_1  -  \varphi_2 \right) \sin \theta_1  \sin \theta_2  +    \\
& \qquad \; \left.   + 8 \left(-2 \lambda_0 + \lambda_- + \lambda_+\right)^2 \cos\left( \varphi_1  -  \varphi_2 \right) \sin\left(2  \theta_1 \right) \sin\left(2  \theta_2 \right)\right\}
.
\end{array}
\label{e-2009-gtq-e3gen}
\end{equation}
For the sake of comparison, let us relate the rather lengthy
correlation coefficient in Eq.~(\ref{e-2009-gtq-e3gen})
to the standard  quantum mechanical correlations based on the dichotomic outcomes
by  setting $\lambda_0 = 0$, $  \lambda_+ = +1$ and  $\lambda_- =-1$.
With these identifications,
\begin{equation}
E_{{ \Psi_{3,2,1}}\,-1, 0, +1 } ({\hat \theta},{\hat \varphi} )= -\frac{2}{3}
\left[\cos \theta_1 \cos \theta_2 + \cos (\varphi_1 - \varphi_2) \sin \theta_1 \sin \theta_2\right]
= \frac{2}{3}E_{{ \Psi_{2,2,1}}\,-1, +1 } ({\hat \theta},{\hat \varphi} )
\label{2009-gtq-edosgc3}
.
\end{equation}
This correlation coefficient is functionally identical with the spin one-half (two outcomes) correlation coefficients.

The correlation coefficient resulting from the Kochen-Specker observable corresponding to
$\lambda_+ = \lambda_- = 1$ and $\lambda_0 = 0$ or its inverted form
$\lambda_+ = \lambda_- = 0$ and $\lambda_0 = 1$
is
\begin{equation}
\begin{array}{rcl}
E_{{ \Psi_{3,2,1}}\,+1, 0, +1 } ({\hat \theta},{\hat \varphi} )&=&
\frac{1}{24} \left\{
11
+\cos [2 (\varphi_1-\varphi_2)]
+4 \cos (\varphi_1-\varphi_2) \sin (2  \theta_1 ) \sin (2  \theta_2 ) +  \right.
\\
&&\;
+2 \left[\cos (2  \theta_1 )+\cos (2 \theta_2) \right] \sin ^2(\varphi_1-\varphi_2)+
\\
&&\; \left.
+\cos (2  \theta_1 )   \cos (2  \theta_2 ) \left[ \cos (2 (\varphi_1-\varphi_2))+3\right]\right\},
\\
E_{{ \Psi_{3,2,1}}\, 0, +1, 0 } ({\hat \theta},{\hat \varphi} )&=&
\frac{1}{3} \left[\cos \theta_1  \cos \theta_2)+\cos    (\varphi_1-\varphi_2) \sin  \theta_1  \sin  \theta_2 \right]^2 ,
\\
E_{{ \Psi_{3,2,1}}\,+1, 0, +1 } (\frac{\pi}{2},\frac{\pi}{2},{\hat \varphi} )&=&
\frac{1}{6} \left\{\cos \left[2 (\varphi_1-\varphi_2)\right]+3\right\},
\\
E_{{ \Psi_{3,2,1}}\, 0, +1, 0 } (\frac{\pi}{2},\frac{\pi}{2},{\hat \varphi} )&=&
\frac{1}{3} \cos ^2(\varphi_1-\varphi_2),
\\
E_{{ \Psi_{3,2,1}}\,+1, 0, +1 } ({\hat \theta},0,0 )&=&
\frac{1}{6} \left\{\cos \left[2 ( \theta_1 - \theta_2 )\right]+3\right\},
\\
E_{{ \Psi_{3,2,1}}\, 0, +1, 0 } ({\hat \theta},0,0 )&=&
\frac{1}{3} \cos ^2( \theta_1 - \theta_2 ).
\end{array}
\label{2009-gtq-edosgc3ks}
\end{equation}

By comparing the quantum correlation coefficient
$E_{{ \Psi_{3,2,1}}\,-1, 0, +1 } ({\hat \theta},0,0 )\propto - \cos (\theta_1 - \theta_2)$
of the spin operators in Eq.~(\ref{2009-gtq-edosgc3})
with the quantum correlation coefficient  of the Kochen Specker operators
$E_{{ \Psi_{3,2,1}}\,+1, 0, +1 } ({\hat \theta},0,0 ) \propto
\cos \left[2 ( \theta_1 - \theta_2 )\right]$
of Eq.~(\ref{2009-gtq-edosgc3ks}),
one could, for higher-than one-half angular momentum observables, envision an ``enhancement'' of the quantum correlation coefficient
by adding weighted correlation coefficients, generated from different labels $\lambda_i$.
Indeed, in the domain $0< | \theta_1 -\theta_2 | < \frac{\pi }{3}$,
the plasticity of
$E_{{ \Psi_{l,2,1}}\,\lambda_{-1},  \lambda_{0} , \lambda_{+1} }$
can be used to build up ``enhanced'' quantum correlations {\it via}
\begin{equation}
\begin{array}{rcl}
&&\frac{1}{2}\left\{
E_{{ \Psi_{3,2,1}}\,-1, 0, +1 } ({\hat \theta},0,0 )
+
3\left[2 E_{{ \Psi_{3,2,1}}\,+1, 0, +1 } ({\hat \theta},0,0 ) -1\right]\right\}
\\
&&\qquad =  \frac{1}{2}\left[-\cos (\theta_1 -\theta_2 ) + \cos 2 (\theta_1 -\theta_2 )
\right]
\\
&& \qquad \qquad
< -\cos (\theta_1 -\theta_2 ) =  E_{{ \Psi_{2,2,1}}\,-1, +1 } ({\hat \theta},0,0 )
\end{array}
\label{2009-gtq-eqcs3}
\end{equation}


\subsection{Four-state particles}

\subsubsection*{Observables}
The angular momentum operator in arbitrary direction $\theta$, $\varphi$ can be written in its spectral form
\begin{equation}
S_\frac{3}{2} (\theta ,\varphi) =-\frac{3}{2}F_{-\frac{3}{2}}(\theta ,\varphi) - \frac{1}{2} F_{-\frac{1}{2}}(\theta ,\varphi) +
\frac{1}{2}F_{+\frac{1}{2}}(\theta ,\varphi)+ \frac{3}{2}F_{+\frac{3}{2}}(\theta ,\varphi).
\label{e-2009-gtq-s444}
\end{equation}

If one is only interested in spin state measurements with the associated outcomes of spin states in units of $\hbar$,
the associated two-particle operator is given by    $
S_{\frac{3}{2}}( \theta_1,\varphi_1 )
\otimes
S_{\frac{3}{2}}( \theta_2,\varphi_2 )$.
More generally, one could define a two-particle operator by
\begin{equation}
F^2_{\lambda_{-\frac{3}{2}},\lambda_{-\frac{1}{2}}, \lambda_{+\frac{1}{2}}, \lambda_{+\frac{3}{2}} } ({\hat \theta},{\hat \varphi} ) =
F_{\lambda_{-\frac{3}{2}},\lambda_{-\frac{1}{2}}, \lambda_{+\frac{1}{2}}, \lambda_{+\frac{3}{2}} } ( \theta_1,\varphi_1)
\otimes
F_{\lambda_{-\frac{3}{2}},\lambda_{-\frac{1}{2}}, \lambda_{+\frac{1}{2}}, \lambda_{+\frac{3}{2}} } ( \theta_2,\varphi_2 ),
\label{2004-gtq-e3F2gen}
\end{equation}
where
\begin{equation}
F_{\lambda_{-\frac{3}{2}},\lambda_{-\frac{1}{2}}, \lambda_{+\frac{1}{2}}, \lambda_{+\frac{3}{2}} } ( \theta ,\varphi )
=
 \lambda_{-\frac{3}{2}}F_{-\frac{3}{2}}(\theta ,\varphi) +  \lambda_{-\frac{1}{2}} F_{-\frac{1}{2}}(\theta ,\varphi) +
 \lambda_{\frac{1}{2}}F_{+\frac{1}{2}}(\theta ,\varphi)+  \lambda_{\frac{3}{2}}F_{+\frac{3}{2}}(\theta ,\varphi)
.
\label{2004-gtq-e3F2gen2}
\end{equation}
For the sake of the physical interpretation of this operator~(\ref{2004-gtq-e3F2gen}), let us consider as a concrete example
a spin state measurement on two quanta:
$ F_{\lambda_{-\frac{3}{2}}}(\theta_1 ,\varphi_1)\otimes   F_{\lambda_{+\frac{3}{2}}}(\theta_2 ,\varphi_2 )$  stands for the proposition
\begin{quote}
{\em `The outcome of the first particle measured along $\theta_1,\varphi_1$ is ``$\lambda_{-\frac{3}{2}}$''
      and
      the outcome of the second particle measured along $\theta_2,\varphi_2$ is ``$\lambda_{+\frac{3}{2}}$''~.'
}
\end{quote}

\subsubsection*{Singlet state}

The singlet state of two spin-$3/2$ observables
can be found by the general methods developed in Ref.~\cite{schimpf-svozil}.
In this case, this amounts to summing all possible two-partite states yielding zero angular momentum,
multiplied with the corresponding  Clebsch-Gordan coefficients
$
\langle j_1m_1j_2m_2\vert 00\rangle = \delta_{j_1,j_2}  \delta_{m_1,-m_2} \frac{(-1)^{j_1-m_1}}{\sqrt{2j_1+1}}
$
of mutually negative single particle states resulting in total angular momentum zero.
More explicitly,  for $j_1=j_2=\frac{3}{2}$,
$
\left|  \left. \psi_{4,2,1} \right\rangle  \right. =
\frac{1}{2} \left(
\left| \left. \frac{3}{2}, -\frac{3}{2}\right\rangle \right.
 - \left| \left.  -\frac{3}{2}, \frac{3}{2}\right\rangle    \right.
- \left| \left.  \frac{1}{2}, -\frac{1}{2}\right\rangle  \right.
+ \left| \left.  -\frac{1}{2}, \frac{1}{2}\right\rangle   \right.
\right)$.
Again, this two-partite singlet state satisfies the uniqueness property.
The four different spin states can be identified with the Cartesian basis of four-dimensional Hilbert space
$\left| \left. \frac{3}{2}\right\rangle \right. \equiv (1,0,0,0)$,
$\left| \left. \frac{1}{2}\right\rangle \right. \equiv (0,1,0,0)$,
$\left| \left. -\frac{1}{2}\right\rangle \right. \equiv (0,0,1,0)$,
and
$\left| \left. -\frac{3}{2}\right\rangle \right. \equiv (0,0,0,1)$,
respectively.

\subsubsection*{Results}

For the sake of comparison, let us again specify the rather lengthy
correlation coefficient in the case of general observables with arbitrary outcomes $\lambda_i$, $i=1,\ldots ,4$
to the standard  quantum mechanical correlations~(\ref{2009-gtq-edosgc3})
by setting $\lambda_{+\frac{3}{2}} = +\frac{3}{2}$,
$\lambda_{+\frac{1}{2}}= +\frac{1}{2}$,
$\lambda_{-\frac{1}{2}}=-\frac{1}{2}$ and
$\lambda_{-\frac{3}{2}}=-\frac{3}{2}$; i.e., by substituting the general outcomes with spin state observables in units of $\hbar$.
With these identifications, the correlation coefficients can be directly calculated {\it via} $S_{\frac{3}{2}\frac{3}{2}}$; i.e.,
\begin{equation}
\begin{array}{rcl}
E_{{ \Psi_{4,2,1}}\,-\frac{3}{2},-\frac{1}{2}, +\frac{1}{2}, +\frac{3}{2} } ({\hat \theta},{\hat \varphi} )
&=&
{\rm Tr}\left\{ \rho_{ \Psi_{4,2,1}} \cdot \left[ S_{\frac{3}{2}}(\theta_1,\varphi_1) \otimes S_{\frac{3}{2}}(\theta_2,\varphi_2)\right]\right\} \\
&=& -\frac{5}{4} \left[\cos \theta_1 \cos \theta_2 + \cos (\varphi_1 - \varphi_2) \sin \theta_1 \sin \theta_2\right] \\
&=& \frac{8}{15}E_{{ \Psi_{2,3,1}}\,-1, +1 } ({\hat \theta},{\hat \varphi} ) \\
&=& 5 E_{{ \Psi_{2,2,1}}\,-\frac{1}{2}, +\frac{1}{2} } ({\hat \theta},{\hat \varphi} )
= \frac{5}{4}E_{{ \Psi_{2,2,1}}\,-1, +1 } ({\hat \theta},{\hat \varphi} )
\end{array}
\label{2009-gtq-edosgc4}
.
\end{equation}
This correlation coefficient is again functionally identical with the spin one-half and spin one (two and three outcomes) correlation coefficients.

\begin{figure}[htbp]
  \centering
\includegraphics[width=60mm]{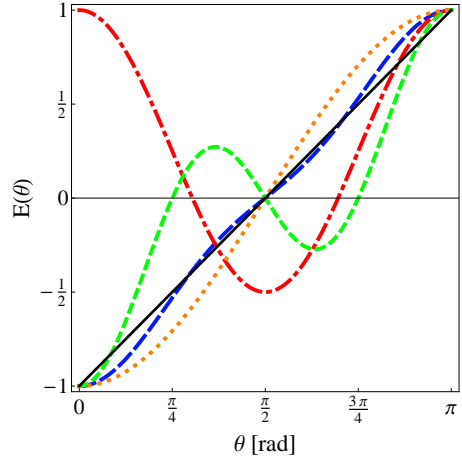}
\caption{Plasticity of the correlation coefficient of two spin three-half quanta in a singlet state.
(a) $E_{{ \Psi_{4,2,1}}\,-1,-1, +1, +1 }$ is represented by the long-dashed blue curve,
(b) $E_{{ \Psi_{4,2,1}}\,-1,+1, +1, -1  }$ is represented by the dashed-dotted red curve,
(c) $E_{{ \Psi_{4,2,1}}\,+1,-1, +1, -1  }$ is represented by the short-dashed green curve,
(d) $\frac{4}{5}E_{{ \Psi_{4,2,1}}\,-\frac{3}{2},-\frac{1}{2}, +\frac{1}{2}, +\frac{3}{2} }$ is represented by the dotted orange curve,
and (e) $E_{\text{cl},2,2}(\theta )$ is represented by the classical linear  black line.
}
\label{2009-gtq-gr4}
\end{figure}

The plasticity of the general correlation coefficient
\begin{equation}
E_{{ \Psi_{4,2,1}}\,\lambda_{-\frac{3}{2}},\lambda_{-\frac{1}{2}}, \lambda_{+\frac{1}{2}}, \lambda_{+\frac{3}{2}} } ({\hat \theta},{\hat \varphi} )
=
{\rm Tr}\left[ \rho_{ \Psi_{4,2,1}} \cdot  F^2_{\lambda_{-\frac{3}{2}},\lambda_{-\frac{1}{2}}, \lambda_{+\frac{1}{2}}, \lambda_{+\frac{3}{2}} } ({\hat \theta},{\hat \varphi})\right]
\label{2009-gtq-gpe4}
\end{equation}
can be demonstrated by enumerating special cases; e.g.,
\begin{equation}
\begin{array}{rcl}
E_{{ \Psi_{4,2,1}}\,-1,-1, +1, +1 } ( \theta ,0,0,0 )
&=& \frac{1}{8} \left[-7 \cos \theta -\cos (3 \theta )\right]
,
\\
E_{{ \Psi_{4,2,1}}\,-1,+1, +1, -1 } ( \theta ,0,0,0 )
&=& \frac{1}{4} \left[3 \cos (2 \theta )+1\right]
,
\\
E_{{ \Psi_{4,2,1}}\,+1,-1, +1, -1 } ( \theta ,0,0,0 )
&=& \frac{1}{2} \left[-\cos \theta -\cos (3 \theta )\right]
.
\\
\end{array}
\label{2009-gtq-e4-plast}
\end{equation}
These functions are drawn in Fig.~\ref{2009-gtq-gr4},
together with the spin state correlation coefficient
$\frac{4}{5}E_{{ \Psi_{4,2,1}}\,-\frac{3}{2},-\frac{1}{2}, +\frac{1}{2}, +\frac{3}{2} }
( \theta ,0,0,0 ) = -\cos \theta $
and the classical linear correlation coefficient
$E_{\text{cl},2,2}(\theta ) = {2 \theta / \pi} - 1$.


\subsection{General case of two spin $j$ particles}

The general case of spin correlation values of two particles with arbitrary spin $j$
(see the Appendix of Ref.~\cite{svozil-krenn} for a group theoretic derivation) can be directly calculated in an analogous way as before, yielding
\begin{equation}
\begin{array}{rcl}
E_{{ \Psi_{2,2j+1,1}}\,-j,-j+1,\ldots,+j-1, +j } ({\hat \theta},{\hat \varphi} )
&=&
{\rm Tr}\left\{ \rho_{ \Psi_{2,2j+1,1}} \cdot \left[ S_{j}(\theta_1,\varphi_1) \otimes S_{j}(\theta_2,\varphi_2)\right]\right\} \\
&=& -\frac{j(1+j)}{3} \left[\cos \theta_1 \cos \theta_2 + \cos (\varphi_1 - \varphi_2) \sin \theta_1 \sin \theta_2\right]  .
\end{array}
\label{2009-gtq-edosgcjj}
\end{equation}
Thus, the functional form of the two-particle correlation coefficients based on spin state observables is {\em
independent} of the absolute spin value.


\section{Four spin one-half particle correlations}

To begin with the analysis of four-partite correlations,  consider four spin-${1\over 2}$
particles in one of the two singlet states~\cite{schimpf-svozil}
$
\vert \Psi_{2,4,1} \rangle
=
{1\over \sqrt{3}}\Bigl[
\vert ++-- \rangle +
\vert --++ \rangle \nonumber
-  {1\over 2}
\bigl(
\vert +- \rangle +
\vert -+ \rangle
\bigr)
\bigl(
\vert +- \rangle +
\vert -+ \rangle
\bigr)
\Bigr],
$
and
$
\vert \Psi_{2,4,2} \rangle
=
\left( \vert \Psi_{2,2,1} \rangle \right)^2
=
{1\over 2}
\bigl(
\vert +- \rangle -
\vert -+ \rangle
\bigr)
\bigl(
\vert +- \rangle -
\vert -+ \rangle
\bigr)$,
where
$\vert \Psi_{2,2,1} \rangle = \frac{1}{\sqrt{ 2}}
\bigl(
\vert +- \rangle -
\vert -+ \rangle
\bigr)
$
is the two particle singlet ``Bell'' state.
In what follows, we shall concentrate on the first state
$\vert \Psi_{2,4,1} \rangle$, since  $\vert \Psi_{2,4,2} \rangle$
is just the product of two two-partite singlet states,
thus presenting entanglement merely among two pairs of two quanta.


The projection operators $F$
corresponding to a four spin one-half particle joint measurement
aligned (``$+$'') or antialigned  (``$-$'') along those angles are
\begin{equation}
\begin{array}{lll}
 F_{\pm \pm \pm \pm} ({\hat \theta},{\hat \varphi} ) =
{\frac{1}{2}}\left[{\mathbb I}_2 \pm {\bf \sigma}( \theta_1,\varphi_1 )\right]
\otimes
{\frac{1}{2}}\left[{\mathbb I}_2 \pm {\bf \sigma}( \theta_2,\varphi_2 )\right] \otimes
\nonumber\\
\qquad\qquad\qquad\qquad\qquad
\otimes
{\frac{1}{2}}\left[{\mathbb I}_2 \pm {\bf \sigma}( \theta_3,\varphi_3 )\right]
\otimes
{\frac{1}{2}}\left[{\mathbb I}_2 \pm {\bf \sigma}( \theta_4,\varphi_4 )\right].
\end{array}
\label{2004-gtq-e2}
\end{equation}

To demonstrate its physical interpretation, let us consider a concrete example: $F_{- + - + } ({\hat \theta},{\hat \varphi} )$ stands for the proposition
\begin{quote}
{\em `The spin state of the first particle measured along $\theta_1,\varphi_1$ is ``$-$'',
      the spin state of the second particle measured along $\theta_2,\varphi_2$ is ``$+$'',
      the spin state of the third particle measured along $\theta_3,\varphi_3$ is ``$-$'',
      and the spin state of the fourth particle measured along $\theta_4,\varphi_4$ is ``$+$''~.'
}
\end{quote}


The joint probability to register the spins of the four particles
in state $\Psi_{2,4,1}$
aligned or anti-aligned along the directions defined by
($\theta_1$, $\varphi_1 $),
($\theta_2$, $\varphi_2 $),
($\theta_3$, $\varphi_3 $),  and
($\theta_4$, $\varphi_4 $) can be evaluated by a straightforward calculation
of
\begin{equation}
P_{{\Psi_{2,4,1}} \pm 1 ,\pm 1,\pm 1\pm 1} ({\hat \theta},{\hat \varphi} )=
{\rm Tr}\left[\rho_{\Psi_{2,4,1}} \cdot F_{\pm \pm \pm \pm} \left({\hat \theta},{\hat \varphi} \right)\right].
\end{equation}

\begin{figure}[htbp]
  \centering
\includegraphics[width=60mm]{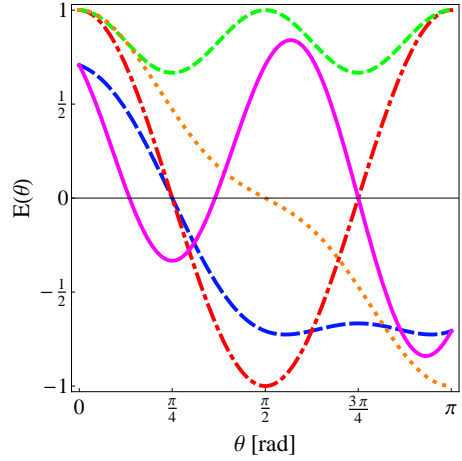}
\caption{Plasticity of the correlation coefficient of four spin one-half quanta in a singlet state.
(a) $E_{{\Psi_{2,4,1}}-1,+1}(\theta , \frac{\pi}{4}, -\theta , \theta )$ is represented by the long-dashed blue curve,
(b) $E_{{\Psi_{2,4,1}}-1,+1}(\theta , \theta , -\theta , \theta )$ is represented by the dashed-dotted red curve,
(c) $E_{{\Psi_{2,4,1}}-1,+1}(\theta , -\theta , -\theta , \theta )$ is represented by the short-dashed green curve,
(d) $E_{{\Psi_{2,4,1}}-1,+1}(\theta , -\theta , -\theta , 0)$ is represented by the dotted orange curve,
and
(e) $E_{{\Psi_{2,4,1}}-1,+1}(-\theta , -\theta , \frac{\pi}{4}, \theta )$ is represented by the solid magenta line.
}
\label{2009-gtq-E241}
\end{figure}

The correlation coefficients and joint probabilities to find the four particles
in an even or in an odd number of
spin-``$-$''-states when measured along
($\theta_1$, $\varphi_1 $),
($\theta_2$, $\varphi_2 $),
($\theta_3$, $\varphi_3 $),  and
($\theta_4$, $\varphi_4 $)
obey  $P_{ \rm even} + P_{ \rm odd}=1$,
as well as $E= P_{ \rm even} - P_{ \rm odd}$; hence
$
P_{ \rm even} =
{1\over2}\left[1 + E  \right]
$
and
$
P_{\rm odd} =
{1\over2}\left[1 - E  \right]
$.
Thus, the four particle quantum correlation is given by (cf. Table~\ref{2008-gtq-2part})
\begin{equation}
\begin{array}{rcl}
E_{ \Psi_{2,4,1}-1,+1}({\hat \theta} , {\hat \varphi } )  &=&
\frac{1}{3}
\left\{
\cos \theta_3 \sin \theta_1
\left[
-\cos \theta_4 \cos (\varphi_1 - \varphi_2) \sin \theta_2 +
          2 \cos \theta_2 \cos (\varphi_1 - \varphi_4) \sin \theta_4
\right] +
\right.
\\
&&\qquad
    \sin \theta_1 \sin \theta_3
\left[2 \cos \theta_2 \cos \theta_4 \cos (\varphi_1 - \varphi_3)  +
\right.
\\
&&\qquad
\qquad
\left.
\left(
2 \cos (\varphi_1 + \varphi_2 - \varphi_3 - \varphi_4) +
                \cos (\varphi_1 - \varphi_2)
                \cos (\varphi_3 - \varphi_4)
\right) \sin \theta_2 \sin \theta_4
\right]   +
\\
&&\qquad
    \cos \theta_1
\left[
2 \sin \theta_2
\left(
\cos \theta_4 \cos (\varphi_2 - \varphi_3) \sin \theta_3 +
                \cos \theta_3 \cos (\varphi_2 - \varphi_4) \sin \theta_4
\right) \right.
 +
\\
&&\qquad
\qquad
\left.
\left.
\cos \theta_2
\left(3 \cos \theta_3 \cos \theta_4 -
                \cos (\varphi_3 - \varphi_4) \sin \theta_3
\sin \theta_4
\right)
\right]
\right\} .
\end{array}
\label{2009-gtq-fpqcgen}
\end{equation}
If all the polar angles $\hat \theta$ are set to $\pi /2$,
then this correlation function yields
\begin{equation}
E_{{\Psi_{2,4,1}}-1,+1} ( \frac{\pi}{2},\frac{\pi}{2},\frac{\pi}{2},\frac{\pi}{2},\hat \varphi )=
\frac{1}{3} \left[2 \cos (\varphi_1+\varphi_2- \varphi_3 - \varphi_4)
+\cos (\varphi_1-\varphi_2) \cos (\varphi_3-\varphi_4)
\right]
.
\end{equation}
Likewise, if all the azimuthal angles $\hat \varphi$ are all set to zero, one obtains
\begin{equation}
E_{{\Psi_{2,4,1}}-1,+1} (\hat \theta )=
\frac{1}{3} \left[2 \cos (\theta_1+\theta_2- \theta_3 - \theta_4)
+\cos (\theta_1-\theta_2) \cos (\theta_3-\theta_4)
\right]
.
\label{2009-gtq-E241e}
\end{equation}
The plasticity of the correlation coefficient
$E_{{\Psi_{2,4,1}}-1,+1} (\hat \theta )$ of Eq.~(\ref{2009-gtq-E241e})
for various parameter values $\theta$ is depicted in Fig.~\ref{2009-gtq-E241}.

\begin{table}
\begin{tabular}{c}
\hline\hline
$
P_{ \rm even} =
{1\over2}\left[1 + E  \right]
\; ,\;
P_{\rm odd} =
{1\over2}\left[1 - E  \right]
\; ,\;
E=
P_{{\rm even}}
-
P_{{\rm odd}}
$
\\
\hline
$
\begin{array}{lll}
E_{{\Psi_{2,4,1}}-1,+1} ({\hat \theta} ,{\hat \varphi})  &=&
\frac{1}{3}
\left\{
\cos \theta_3 \sin \theta_1
\left[
-\cos \theta_4 \cos (\varphi_1 - \varphi_2) \sin \theta_2 +
          2 \cos \theta_2 \cos (\varphi_1 - \varphi_4) \sin \theta_4
\right] +
\right.
\\
&&\qquad
    \sin \theta_1 \sin \theta_3
\left[2 \cos \theta_2 \cos \theta_4 \cos (\varphi_1 - \varphi_3)  +
\right.
\\
&&\qquad
\qquad
\left.
\left(
2 \cos (\varphi_1 + \varphi_2 - \varphi_3 - \varphi_4) +
                \cos (\varphi_1 - \varphi_2)
                \cos (\varphi_3 - \varphi_4)
\right) \sin \theta_2 \sin \theta_4
\right]   +
\\
&&\qquad
    \cos \theta_1
\left[
2 \sin \theta_2
\left(
\cos \theta_4 \cos (\varphi_2 - \varphi_3) \sin \theta_3 +
                \cos \theta_3 \cos (\varphi_2 - \varphi_4) \sin \theta_4
\right) \right.
 +
\\
&&\qquad
\qquad
\left.
\left.
\cos \theta_2
\left(3 \cos \theta_3 \cos \theta_4 -
                \cos (\varphi_3 - \varphi_4) \sin \theta_3
\sin \theta_4
\right)
\right]
\right\}
\end{array}
$
\\
$E_{{\Psi_{2,4,1}}-1,+1} ({\hat \theta} )  =
\frac{1}{3} \left[2 \cos (\theta_1 +\theta_2 -\theta_3 -\theta_4 )+\cos
   (\theta_1 -\theta_2 ) \cos (\theta_3 -\theta_4 )\right].
$
\\
$
E_{{\Psi_{2,4,1}}-1,+1} ( \frac{\pi}{2},\frac{\pi}{2},\frac{\pi}{2},\frac{\pi}{2},\hat \varphi )=
\frac{1}{3} \left[2 \cos (\varphi_1+\varphi_2- \varphi_3 - \varphi_4)
+\cos (\varphi_1-\varphi_2) \cos (\varphi_3-\varphi_4)
\right]
$ \\
\hline
$
\begin{array}{lll}
E_{{\Psi_{2,4,2}}-1,+1}({\hat \theta} , {\hat \varphi } )  &=&
\left[\cos \theta_1 \cos \theta_2 +
          \cos ( \varphi_1 - \varphi_2) \sin \theta_1 \sin \theta_2\right]\cdot \\
&&\qquad  \qquad  \left[\cos \theta_3 \cos \theta_4 +
          \cos (\varphi_3 - \varphi_4) \sin \theta_3 \sin \theta_4
\right]
\end{array}
$
\\
$E_{{\Psi_{2,4,2}}-1,+1}({\hat \theta} )  =
\cos (\theta_1 -\theta_2 ) \cos (\theta_3 -\theta_4 ),
$
\\
$E_{{\Psi_{2,4,2}}-1,+1}( \frac{\pi}{2},\frac{\pi}{2},\frac{\pi}{2},\frac{\pi}{2},{\hat \varphi} )  =
\cos (\varphi_1 -\varphi_2 ) \cos (\varphi_3 -\varphi_4 ),
$
\\
\hline\hline
\end{tabular}
\caption{Probabilities and correlation coefficients
for finding an odd or even number of spin-``$-$''-states for both four-partite singlet states.
Omitted arguments are zero.
\label{2008-gtq-2part}
}
\end{table}

\section{Summary}

Compared to the two-partite quantum correlations of two-state particles,
the plasticity of the quantum correlations of states of {\em more than  two particles }
originates in the dependency of the {\em multitude of angles} involved, as well as by the {\em multitude of singlet states} in this domain.
For states composed from particles of {\em more than two mutually exclusive outcomes,} the plasticity
is also increased by the {\em different values associated with the outcomes.}

We have explicitly derived the quantum correlation functions of two- and four-partite spin one-half, a well as two-partite systems of higher spin.
All quantum correlation coefficients of the two-partite spin observables have identical form, all being proportional to
$\cos \theta_1 \cos \theta_2 + \cos (\varphi_1 - \varphi_2) \sin \theta_1 \sin \theta_2$.
We have also argued that, by utilizing the plasticity of the quantum correlation coefficients for spins higher that one-half,
this well-known correlation function can be ``enhanced'' by defining sums of quantum correlation coefficients,
at least in some domains of the measurement angles.

It would be interesting to know whether this plasticity of the quantum correlations
$E_{{ \Psi_{l,2,1}}\,\lambda_{-l}, \ldots , \lambda_{+l} }$
for ``very high'' angular momentum $l$ observables
could be pushed to the point of maximal violation of
the Clauser-Horne-Shimony-Holt inequality {\em without} a  bit exchange
such as by using the
``buildup'' of a step function from the individual correlation coefficients~\cite{svozil-krenn}; e.g., for  $0 \leq \theta \leq \pi $,
\begin{equation}
{\rm sgn} (x)
=\left\{ \begin{array}{rl}
-1 \; &\text{  for } 0\le x<\frac{\pi}{2} \\
 0 \; &\text{  for }  x=\frac{\pi}{2}  \\
+1 \; &\text{  for }  \frac{\pi}{2} <  \theta \le \pi
   \end{array} \right.
={4\over \pi }\sum_{n=0}^\infty {(-1)^n\cos \left[(2n+1)\left( \theta +\frac{\pi}{2}\right)\right]\over
2n+1}\quad .
\label{e:10}
\end{equation}
Any such violation of Boole-Bell type ``conditions of possible experience'' beyond the maximal quantum
violations, as for instance
derived by Tsirelson~\cite{cirelson} and generalized in Ref.~\cite{filipp-svo-04-qpoly-prl} not necessarily generalizes
to the multipartite, non dichotomic cases.
Note also that such a strong or even maximal violation of the Boole-Bell type ``conditions of possible experience'' beyond the maximal quantum
violations
needs
not necessarily violate relativistic causality~\cite{popescu-97,popescu-97b},
or be associated with a ``sharpening'' of the angular dependence of the joint occurrence of certain elementary dichotomic outcomes,
such as  ``$++$,'' ``$+-$,'' ``$-+$'' or ``$--$,'' respectively.


%

\end{document}